\documentclass[twocolumn,amsmath,amssymb]{revtex4}
\input epsf
\usepackage{appendix}
\usepackage{rotating}
\usepackage[hang,nooneline]{subfigure}
\usepackage{amsmath,amssymb,graphicx}
\usepackage{algorithmic}
\usepackage{enumerate}
\usepackage{color}

\usepackage[pdftex]{hyperref}
\hypersetup{colorlinks=true,linkcolor=blue,citecolor=blue,urlcolor=blue}

\begin{document}
\title {Spectral and dynamical analysis of a single vortex ring in anisotropic harmonically trapped three-dimensional Bose-Einstein condensates}
\author{Christopher Ticknor}
\affiliation{Theoretical Division, Los Alamos National Laboratory, Los Alamos, New Mexico 87545, USA}

\author{Wenlong Wang}
\email{wenlongcmp@gmail.com}
\affiliation{Department of Theoretical Physics, Royal Institute of Technology, Stockholm, SE-106 91, Sweden}

\author{P. G. Kevrekidis}
\affiliation{Department of Mathematics and Statistics, University of Massachusetts,
Amherst, Massachusetts 01003-4515 USA}

\date{\today}
\begin{abstract}
In the present work, motivated by numerous recent experimental developments we revisit the dynamics of a single vortex ring in anisotropic harmonic traps. 
At the theoretical level, we start from a general Lagrangian dynamically capturing the evolution of a vortex ring and not only consider
its spectrum of linearized excitations, but also explore
the full nonlinear dynamical evolution of the ring as a vortical filament. The theory predicts that the ring is stable for $1 \leq \lambda \leq 2$, where $\lambda=\omega_z/\omega_r$ is the ratio of the trapping frequencies along the $z$ and $r$ axes, i.e., for spherical to slightly oblate condensates. We compare this
prediction with direct numerical simulations of the full 3D Gross-Pitaevskii equation (GPE) capturing the linearization spectrum of the ring for different values of the chemical potential and as a function of the anisotropy
parameter $\lambda$. We identify this result as being only asymptotically valid as the chemical potential $\mu \rightarrow \infty$, revealing  how the stability interval narrows and, in particular, its upper bound decreases for
finite $\mu$. Finally, we compare at the dynamical level the results of the GPE with the ones effectively capturing the ring dynamics, revealing the unstable evolution for different values of $\lambda$, as well as the good
agreement between the two.

\end{abstract} 
\maketitle
\section{Introduction}

The study of vortex rings (VRs) has a long and distinguished history
in the context of fluids and superfluids~\cite{saffman,Pismen1999},
a prominent chapter of which involves, e.g., associated observations
in helium; see, for instance~\cite{donnelly,Rayfield64,Gamota73}.
However, the experimental realization and
subsequent developments in the theme of Bose-Einstein
condensates (BECs)~\cite{book1,book2,dumitr1,dumitr2} have produced a
playground especially suitable for the study of such coherent
structures. This is also evidenced by the plethora of reviews and
book chapters on the topic~\cite{emergent,komineas_rev,barenghi_rev,book_new}.

In the early stages of examination of VRs in the context of BECs,
the emphasis was on finding mechanisms for their creation.
These ranged from the (transverse) instability of planar dark solitons
in 3D BECs~\cite{Anderson01}, to the generation via density
engineering in suitable geometries, as in the work of~\cite{Shomroni09},
or even their spontaneous emergence as byproducts of collision
between symmetric defects in~\cite{Ginsberg05}. They were also
detected in the collision between dark solitons imprinted
in non-one-dimensional
geometries, where they were seen to give rise to unusual dynamical
events such as slingshot events of one of the waves~\cite{sengstock}.
More recently, the significant improvements in the visualization
techniques, not only in the context of BECs~\cite{bis1,bis2},
but also in the case of superfluid helium~\cite{lathrop1,lathrop2},
have enabled a much more detailed ability to explore the dynamics
of vortex rings. In light of that, one of the recent focal points
has been the exploration of the interaction between rings both
theoretically and computationally~\cite{wacks,wang2}, as well as experimentally~\cite{bis3}.
This aspect is especially important not only in its own right, but
also due to its critical role in the theory of quantum turbulence,
another topic of intense recent study~\cite{Tsatsos2016,Navon2016}.
It should also be mentioned in passing that in addition to these
significant developments on the dynamics of vortex rings in bosonic
systems, similar (planar) solitonic and vortical (line and ring)
structures have been recently identified in superfluid Fermi
gases~\cite{zwierlein}.

Our emphasis in the present work is different, going partially back
to the context of the single vortex ring. There
exists an analytical prediction for its spectrum, stemming from
the early work of~\cite{horng}. The relevant prediction relates
the potential stability of the VR with the potential prolateness
or oblateness of the full condensate. In particular, defining
$\lambda= \omega_z/\omega_r$, i.e., the ratio of confining frequencies
along the $z$- vs. along the radial-direction, the analysis
of~\cite{horng} yields that the VR is stable when $1 \leq \lambda \leq 2$.
To the best of our knowledge, such a theoretical prediction has not
been systematically tested. Here, we test this prediction, finding that
it is {\it only asymptotically valid} in the regime of $\mu \rightarrow
\infty$. We showcase the intriguing way that the prediction deviates
from its asymptotic form, identifying stable rings for
$1 \leq \lambda \leq \lambda_c(\mu)$, where $\lambda_c$ is an upper critical
bound of stability with $\lim_{\mu \rightarrow \infty} \lambda_c(\mu)=2$ and $\lambda_c$ demonstrating a monotonically decreasing trend, as
the chemical potential $\mu$
decreases within finite values. We also go a step further: we explore
the fully nonlinear dynamical equation for the ring as a vortical
filament, whose linearization results in the above prediction.
This is the full nonlinear equation encompassing the combined dynamics of
the ring and its ``Kelvin modes'', comparing this with the full
GPE dynamics. In our view, this is an
important step towards both a quantitative understanding of the
VR stability, but also a stepping stone towards formulating a
multi-ring Lagrangian and developing
the associated nonlinear dynamics for settings
where this may be relevant (such as, e.g.,~\cite{wacks}).

Our presentation is structured as follows. In Sec.~\ref{theory}, we summarize our theoretical analysis, connecting wherever appropriate
with past theoretical analyses on the subject.
In Sec.~\ref{computations}, we discuss our numerical methods.
Next we explore extensively the comparison of the theoretical findings with the spectral and dynamical results of the full 3D problem in Sec.~\ref{results}. 
We compare the spectra in a wide interval of the anisotropy parameter
$\lambda$ and the chemical potential $\mu$, and we establish the validity of the vortex ring filament 
equation of motion. Additionally,
we also visualize the identified instabilities for different values
of $\lambda$ and for different case examples of excited modes.
This provides us with a reasonably complete sense of the (single)
ring's dynamical properties. This is not only of value
in its own right, but also a necessary preamble towards a more
systematic theoretical understanding of multi-ring configurations.
This and related topics for future study are commented upon, along
with a summary of the present findings in Sec.~\ref{cc}.

\section{Theoretical Analysis}
\label{theory}

The dynamics of vortical filaments is a topic
that was studied rather extensively shortly after vortical
patterns could be produced in BECs. We single out here
two ground-breaking works along this direction, namely~\cite{fetter1}
and~\cite{ruban1}. Following the latter (although equivalent results
for our purposes are obtained in the former), we can write down
the Hamiltonian and the Lagrangian of vortical ring filaments
of radius $R$ and vertical position $Z$ as:
\begin{eqnarray}
H &=& \int_0^{2 \pi} \rho \sqrt{R_\phi^2+ R^2 + Z_\phi^2} d \phi, \\
\label{vr_eq1}
L &=& \int_0^{2 \pi} F Z_t - \rho \sqrt{R_\phi^2+ R^2 + Z_\phi^2} d \phi.
\label{vr_eq2}
\end{eqnarray}
Here, $F(R,Z)$ is such that $F_R=\rho(R,Z) R$, the subscripts
denote partial derivatives with respect to the argument, while
$A=\sqrt{R_\phi^2+ R^2 + Z_\phi^2}$ denotes the arclength quantity
in cylindrical coordinates that are most natural to use in the
present setting. The density $\rho$ is assumed to be well
approximated by the Thomas-Fermi limit (for the large
chemical potential values of interest) in the form
$\rho(R,Z)=\max(\mu-V(R,Z),0)$.
It should be noted that this expression is 
part of a more general formulation for arbitrary
filaments~\cite{fetter1,ruban1}, yet herein we restrict considerations
to a ring geometry and the corresponding natural cylindrical coordinate
choice.

The resulting equations of motion (see also the recent exposition
of~\cite{ruban2}) then read:
\begin{eqnarray}
  \rho R R_t &=& -\rho_Z A + \frac{\partial}{\partial \phi}
  \left(\frac{\rho Z_{\phi}}{A}\right),
  \label{vr_eq3}
  \\
  \rho R Z_t &=& \rho_R A + \frac{\rho}{A} R -  \frac{\partial}{\partial \phi}
  \left(\frac{\rho R_{\phi}}{A}\right).
  \label{vr_eq4}
\end{eqnarray}
To the best of our understanding, these partial differential
equations (PDEs), the canonical description of a vortex ring
as a filamentary structure, have never been explored (even numerically)
in their fully nonlinear form for $R=R(\phi,t)$ and $Z=Z(\phi,t)$.
Instead, only the corresponding ordinary differential equations
(ODEs) have been derived for homogeneous, $\phi$-independent,
steady states, i.e.,:
\begin{eqnarray}
  R_t=\omega_z^2 \frac{Z}{\rho}, \quad
    Z_t=-\omega_r^2 \frac{R}{\rho} + \frac{1}{R}.
    \label{vr_eq5}
\end{eqnarray}
Here, the parabolic potential $V(r,z)=(\omega_r^2 r^2 + \omega_z^2 z^2)/2$
has been assumed. It is important to note that these ODEs are the same
as the ones used by~\cite{horng} to derive the VR spectrum, up to a
logarithmic correction
factor~\cite{fetter1} of:
\begin{eqnarray} 
  \Lambda(r)=-\frac{1}{2} \log \left[\left( \sqrt{\frac{\omega_z^2}{2 \mu}+ \frac{\kappa^2}{8}}
    \right) \frac{1}{\sqrt{2 \mu}} \right].
  \label{vr_eq6}
\end{eqnarray}
Here $\kappa$ denotes the curvature of the filament equal to $1/R$ for
the case of the VR.

Importantly, Eqs.~(\ref{vr_eq5}) lead to an equilibrium ring with
$Z_0=0$ and radius $R_0^2=2\mu/(3 \omega_r^2)=R_{\perp}^2/3$.
 Linearizing around this equilibrium,
at the level of Eqs.~(\ref{vr_eq3})-(\ref{vr_eq4}), by using the
ansatz
\begin{eqnarray}
  Z= \epsilon \sum_m Z_m \cos(m \phi), \quad
  R=R_0 + \epsilon \sum_m R_m \sin(m \phi),
  \label{vr_eq8}
\end{eqnarray}
leads to an effective eigenvalue problem through $R_m=e^{i \omega t} R_m^0$
and $Z_m=e^{i \omega t} Z_m^0$, which can be solved for each $m$. The resulting
expression yields~\cite{horng} for each subspace represented by its
own (integer to ensure periodicity) value of $m$:
\begin{eqnarray}
  \omega = \frac{3\Lambda\omega_r^2}{2\mu} \left[ \left(m^2-\lambda^2\right)
    \left(m^2-3\right) \right]^{1/2}.
 \label{spectrum}
\end{eqnarray}

On the basis of the expression of Eq.~(\ref{spectrum}), it can be seen that
if $\lambda  < 1$ (prolate BECs),
the mode with $m=1$ will always be unstable producing a
negative sign inside the bracket of Eq.~(\ref{spectrum}). In the case where
$1 \leq \lambda \leq 2$, i.e., for slightly oblate BECs, all the modes
of Eq.~(\ref{spectrum}) possess real frequencies and therefore the
VR is expected to be stable. Finally, for $\lambda > 2$, the
the modes with $m=1$ and $m \geq \lambda$ will be stable, while
those ``in between'', i.e., $m=2, \dots$ will necessarily be unstable
yielding at least one unstable mode. For $2< \lambda \leq 3$, there will
only be one such mode, for $3 < \lambda \leq 4$, there will be two,
then three and so on. Moreover, which one of these modes dominates,
in terms of producing the maximal growth rate
depends on $\lambda$, based on the above theory. For instance,
for $3 \leq \lambda < \sqrt{10}$, the mode with $m=2$ dominates
the instability growth rate, while for larger $\lambda$, the
$m=3$ acquires a larger growth rate than the $m=2$ one. Similarly,
at $\lambda=\sqrt{22}$, $m=4$ becomes more unstable than $m=3$,
and more generally two modes $m$ and $\tilde{m}$ exchange their
respective dominance at $\lambda=\sqrt{m^2 + \tilde{m}^2 -3}$.

It is interesting to note that, to the best of our knowledge,
these predictions have not been systematically tested, partly
also due to the intensive nature of the relevant computations.
Such an examination, as a function of $\lambda$, but also varying
$\mu$ will be the main theme of the spectral computations of our
numerical section below. Then, upon appreciating the asymptotic
nature of the relevant comparison, we will also attempt to explore
the unstable dynamics of the VR for $\lambda < 1$, as well as
for $\lambda > 2$, to observe how the corresponding instabilities
of the different modes manifest themselves and what is the comparison
between the predictions of Eqs.~(\ref{vr_eq3})-(\ref{vr_eq4}) and
the original GPE with the three-dimensional parabolic trap.

\section{Computational setup}
\label{computations}

Our numerical simulations include finding stationary states of the vortex ring and computing the corresponding linear stability spectra, 
as well as performing the dynamical integration of both the filament PDE
(i.e., the effective description of Eqs.~(\ref{vr_eq3})-(\ref{vr_eq4}))
and the GPE. We use Newton's iteration method to identify the numerically exact 
VR stationary solutions of the GPE:
\begin{eqnarray}
  i \hbar u_t =-\frac{\hbar^2}{2m} \Delta u + V(r,z) u + g |u|^2 u - \mu u.
  \label{vr_eq9}
\end{eqnarray}
For such stationary solutions, the left hand side of the equation
is set to $0$. Also,
here, $V(r,z)=(\omega_r^2 r^2 + \omega_z^2 z^2)m/2$ is the parabolic
confinement of atoms with mass $m$, while $\Delta$ stands for the 3D Laplacian, 
and $g$ is $4\pi \hbar^2 a_s/m$ with $a_s$ being the s-wave scattering length.
We rescale Eq. (\ref{vr_eq9}) in terms of the energy $\hbar \omega_r$.
This naturally leads to length scales being measured in
oscillator units, i.e., $\sqrt{\hbar/m\omega_r}$ for length,
and time scale in units of the inverse trapping frequency, $1/\omega_r$.
Subsequently, we consider the excitation spectrum around a
solution $u_0(x,y,z)$ of the steady state of Eq.~(\ref{vr_eq9}).
This is done by using the ansatz:
\begin{eqnarray}
  u=u_0(x,y,z)+ a(x,y,z) e^{i \omega t} + b^*(x,y,z) e^{-i \omega^* t}
  \label{vr_eq10}
\end{eqnarray}
The resulting, so-called Bogolyubov-de Gennes (BdG) equations can be used for computing normal mode eigenfrequencies $\omega$ and eigenvectors $(a,b)^T$.
Because a steady ring has rotational symmetry, we compute both the stationary state and the spectrum using a cross section in the $r$-$z$ plane. The spectrum is computed using basis expansions through the so-called partial wave method. 
A recent summary of the technique can be
found, e.g., in \cite{wenlongdss} for one-component solitons with rotational symmetry up to a topological charge and we refer the interested readers to this paper for more details. Nonetheless, we briefly mention here that the method computes eigenvalues for each Fourier $m$-mode separately (eigenvalues of modes $m$ and $-m$ are complex conjugates) and the full 3D spectrum is the union of all the individual 2D spectra. In our work, we have collected the spectrum using $m=0, 1, 2, 3, 4$ and $5$ \cite{wenlongdss}.

The stationary ring states in the $(\lambda,\mu)$ plane are explored using
parametric continuations. 
We take advantage of the linear limit of the ring when $\lambda=2$, where the ring is a superposition of a dark soliton state and 
a dark soliton ring state with a phase difference of $\pi/2$~\cite{tickold}.
The states at larger values of $\mu$ are computed via a parametric
continuation in $\mu$. 
The states at a fixed value of $\mu$ but different $\lambda$ values are then computed by a second parametric continuation in $\lambda$. 
Our fields are all discretized using standard finite element methods,
also for dynamics. We have used two different methods for our dynamics for
comparing the different PDE models. The simpler, effectively
1+1D ($\phi$- and $t$- dependent) dynamics of the
ring filament PDE of (\ref{vr_eq3})-(\ref{vr_eq4})
is simulated using the regular fourth order Runge-Kutta method, while the
more demanding 3+1D GPE  is simulated using a 
third order operator splitting Fourier spectral method.

Finally, we present some details of the initial states used in the dynamical evolution simulations. Here, to emulate more closely 
the experimental protocols customarily used, we imprint (within the BEC) a VR with its center position given by:
\begin{eqnarray}
  R_{\rm{ring}}(\phi)=R_0+\sum_m R_m \sin(m \phi).
  \label{vr_eq11}
\end{eqnarray}
Here, each $m$ represents the excitation of a Kelvin wave
mode of index $m$.
Then in each $r$-$z$ plane, the phase of the VR is given
by $\varphi=\arctan[(z-Z_{\rm{ring}})/(r-R_{\rm{ring}})]$.
Typically, a single $R_m$ is set to 0.1 while all the others are set to zero,
and we have picked $Z_{\rm{ring}}$ to be $10^{-6}$.

\section{Results}
\label{results}

\subsection{Spectra}
\begin{figure}
\includegraphics[width=\columnwidth]{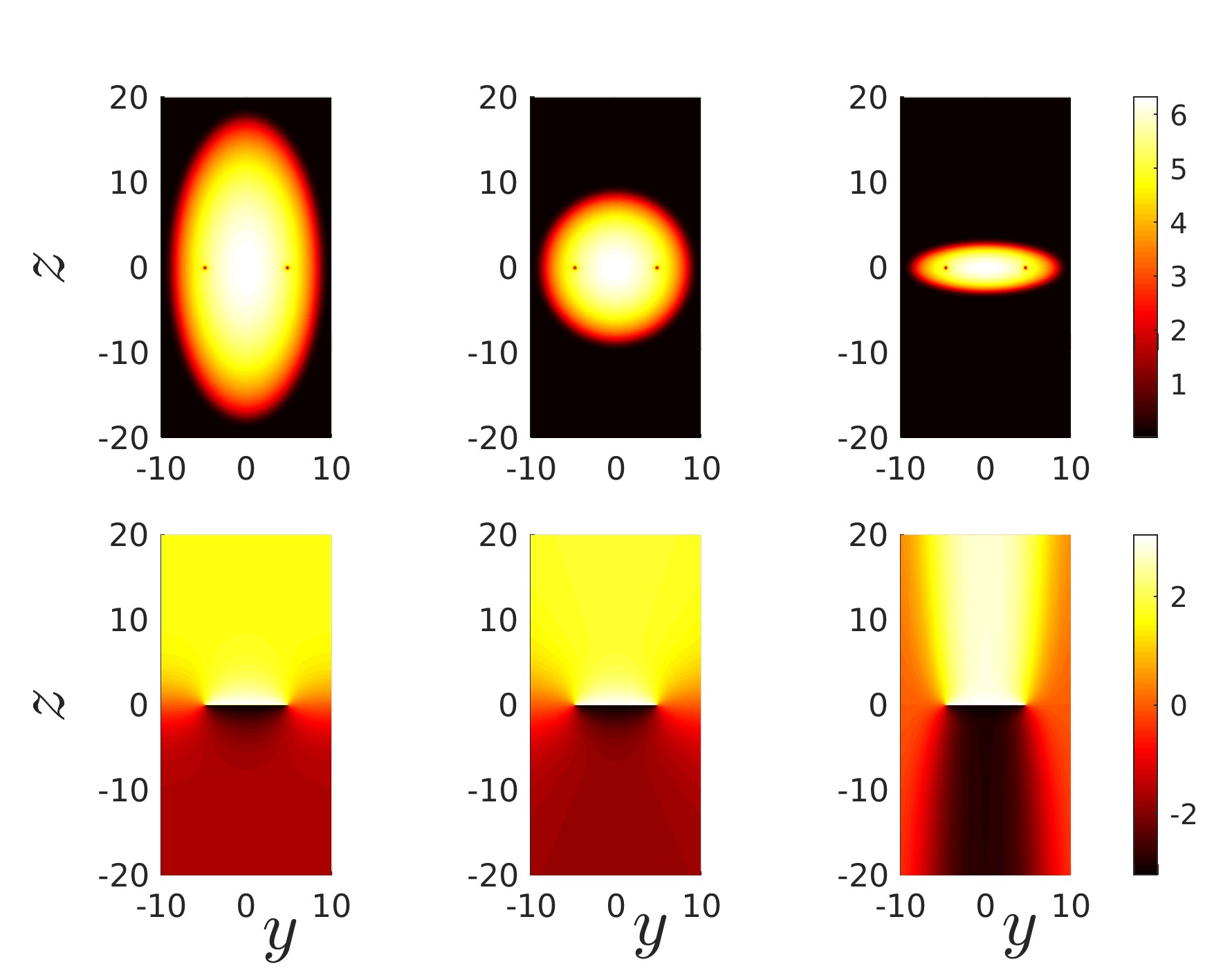}
\put (-225,183) {(a) $\lambda=0.5$}
\put (-145,183) {(b) $\lambda=1$}
\put (-70,183) {(c) $\lambda=3$}
\caption{Three typical states of the VR in various trap geometries in the TF limit at $\mu=40$. 
The figures are for condensates with: (a) prolate, $\lambda=0.5$; (b) spherical, $\lambda=1$;  and (c) oblate, $\lambda=3$ geometries.
The top panels show $|u|$ and the bottom panels illustrate
the phase of the field.
The states have rotational symmetry with respect to the $z$-axis and therefore only a cross section in the $y$-$z$ plane is shown. 
}
\label{states}
\end{figure}

We start by presenting three typical stationary ring states for three different trapping frequencies shown in Fig.~\ref{states}. 
They are from left to right for $\lambda=0.5$ (prolate), $\lambda=1$ (spherical) and $\lambda=3$ (oblate) at $\mu=40$ in the Thomas-Fermi regime. 
Notice how the aspect ratios of the profiles change as we vary
$\lambda$, by changing $\omega_z$. By the way, it is worth mentioning
that since we vary $\omega_z$, the planar ($x$-$y$) profile of the VR remains
that of a dark ring~\cite{tickold}, hence it is not shown here.

\begin{figure*}
\mbox{
\hspace{0cm}
\subfigure[][]{
\includegraphics[width=0.95\columnwidth]{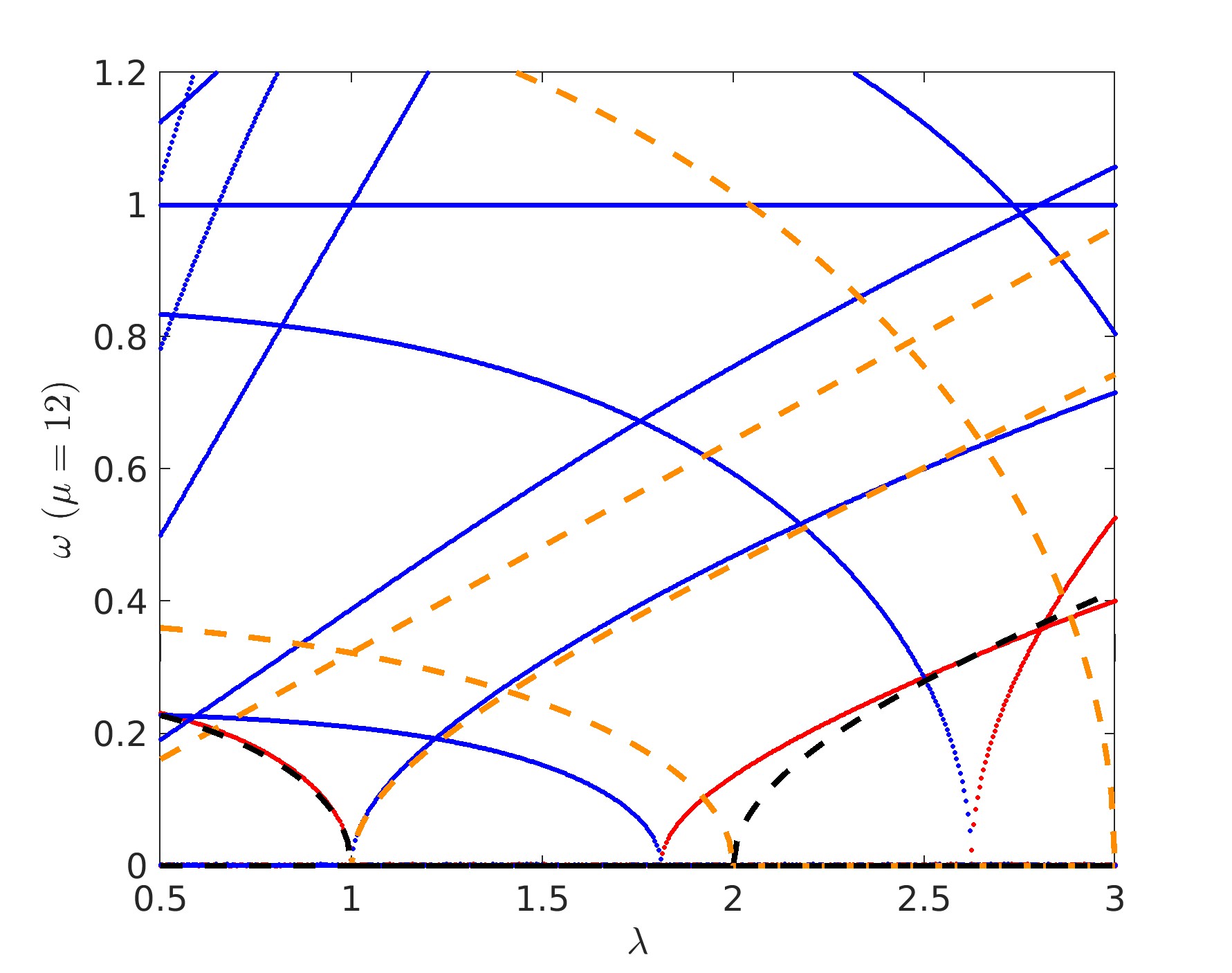}
}
}
\mbox{
\hspace{0cm}
\subfigure[][]{
\includegraphics[width=0.95\columnwidth]{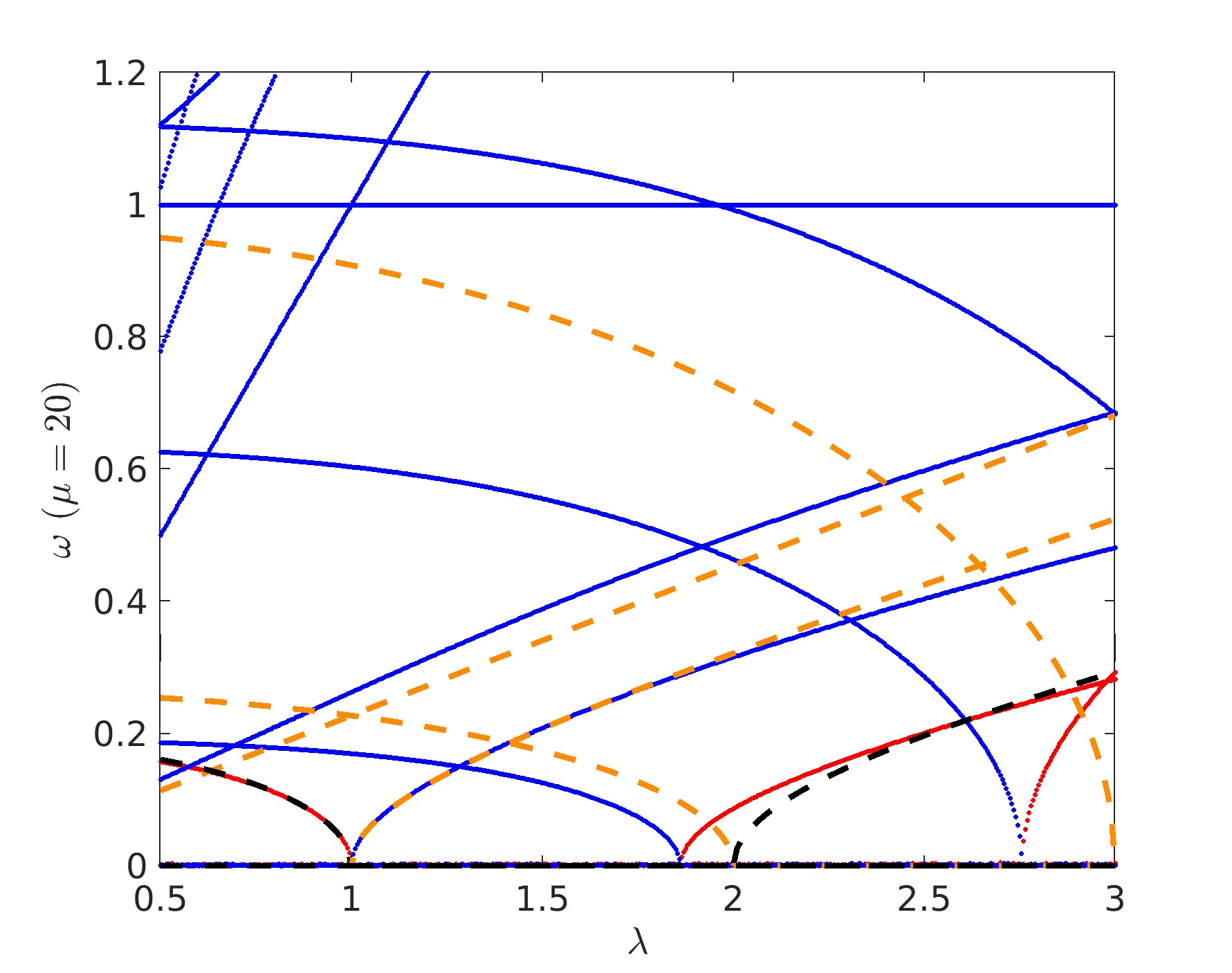}
}
}
\mbox{
\hspace{0cm}
\subfigure[][]{
\includegraphics[width=0.95\columnwidth]{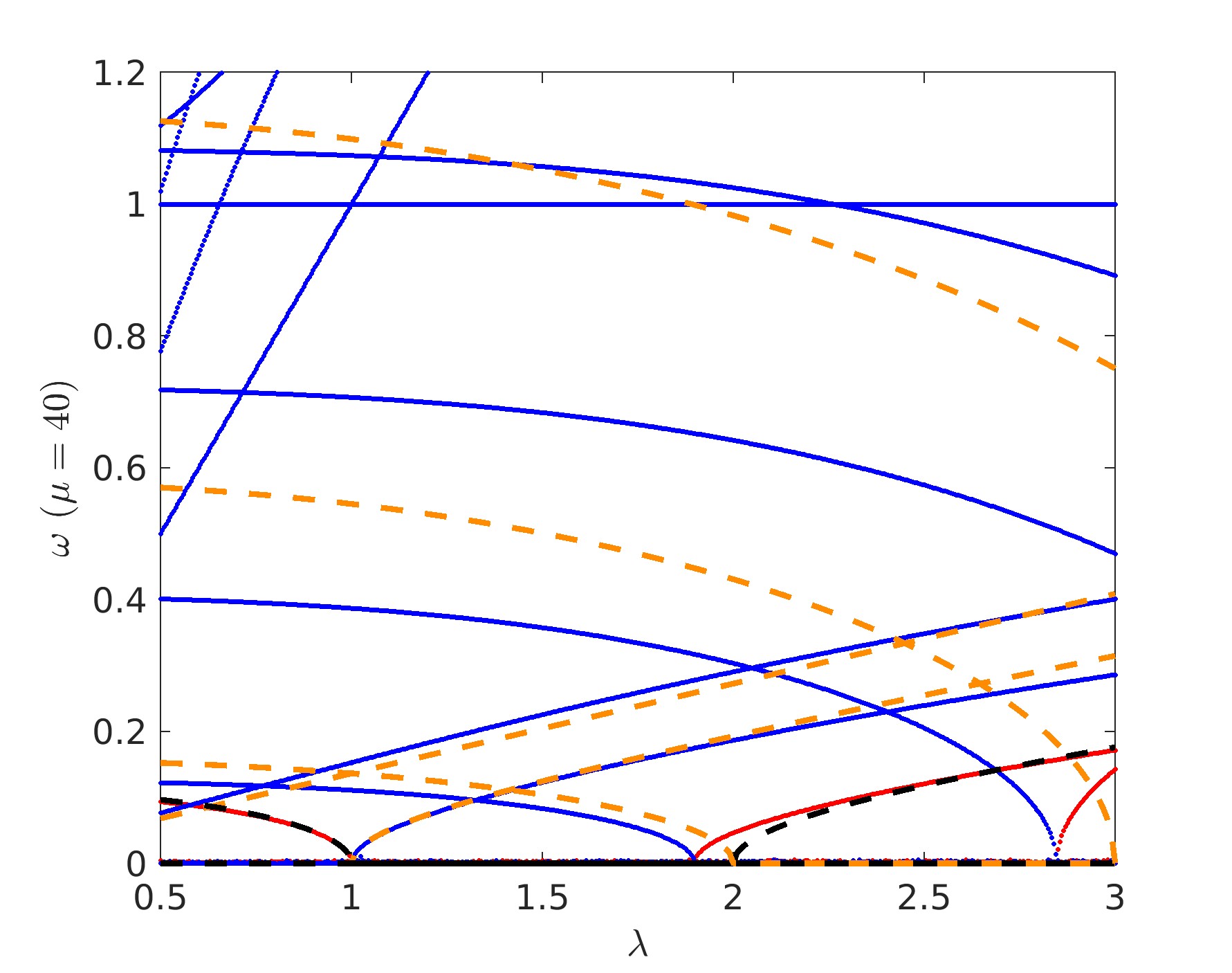}
}
}
\mbox{
\hspace{0cm}
\subfigure[][]{
\includegraphics[width=0.95\columnwidth]{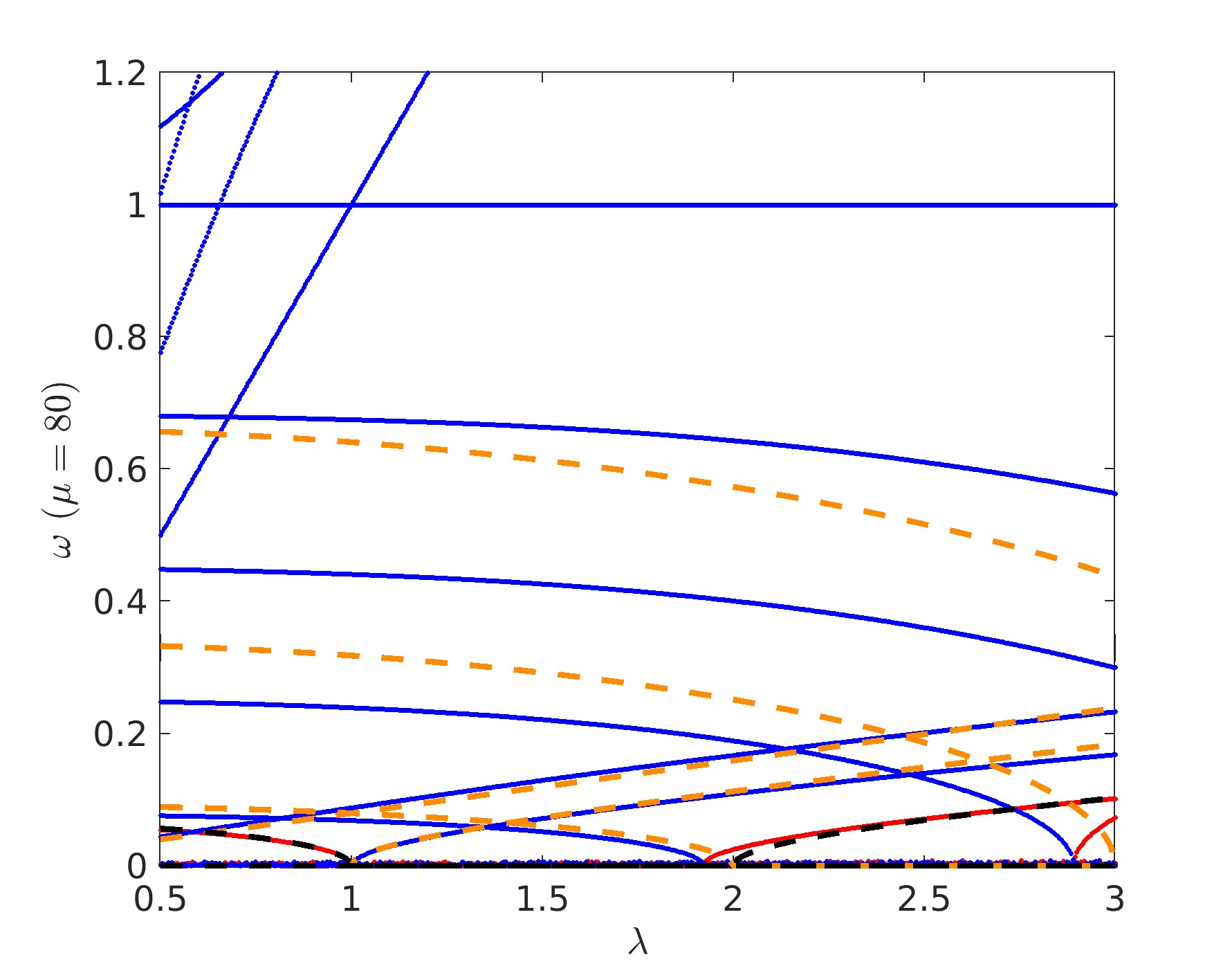}
}
}
\caption{
Four typical spectra of the stationary ring as a function of $\lambda$ for different values of the chemical potential $\mu$.
The blue (stable mode) and red (unstable mode) lines are the full 3D numerical spectra of the GPE, while the gold (stable mode) and black (unstable mode) dashed lines are from the analytical
formula of Eq.~(\ref{spectrum}). Take the analytical lines for example, the one with no instability is the $m=0$ mode, the one that is unstable below $\lambda=1$ is the $m=1$ mode, the one that becomes unstable near $\lambda=2$ is the $m=2$ mode, etc. Note that the two results compare well and become closer as the chemical potential $\mu$ increases. The finite $\mu$ effects are
further discussed in the text.
}
\label{spectra}
\end{figure*}

Next we present four typical results of the spectrum of the stationary ring as a function of $\lambda$ for different values of the 
chemical potential and compare with the analytical results of
Eq.~(\ref{spectrum}) as shown in Fig.~\ref{spectra}.
Notice the good agreement
especially for the lower mode of $m=1$ and the capturing
of the corresponding instability for $\lambda < 1$. Importantly,
note also that this instability is {\it indifferent} to the variation
of the chemical potential and presumably has to do with a change of the
symmetry of the trap from prolate (for $\lambda < 1$) to oblate
(for $\lambda >1$). However, the instability for $m=2$ and higher
does nontrivially depend on the chemical potential $\mu$. Indeed,
as $\mu$ grows, we approach the large $\mu$ limit result of
$\lambda \rightarrow 2$ for the instability threshold. However,
as we depart from that limit, the critical point for the instability,
$\lambda_c$ progressively decreases, lying further
away from $\lambda=2$.

Importantly, this trend of capturing the spectrum better as $\mu \rightarrow
\infty$ is not only evident in the mode with $m=2$, but in fact continues
for higher values of $m$. While the $m=0$ and $m=1$ modes (associated
with ring oscillations inside the trap) are accurately captured more or
less for all values of $\lambda$, the higher modes are progressively
less accurate when $\mu$ is smaller and are becoming more accurate
as $\mu$ increases. This trend has also been observed in a variety of
other problems associated with solitonic and vortical filaments~\cite{ai1}.
In this light, just as the instability of the $m=2$ mode arises for
$\lambda_c<2$, the instability of the $m=3$ mode occurs (for finite
$\mu$) for $\lambda_c < 3$, and so on. Nevertheless, we can see that
Eq.~(\ref{spectrum}) provides an excellent qualitative handle on the nature
of the arising instabilities and the parametric dependence (over $\lambda$)
of the different eigenfrequencies.

\subsection{Dynamics}
Following our spectra of the last section, we now turn to the dynamics. We first illustrate the dynamical effect of
a few key unstable modes predicted by the spectra. 
These results are from full simulations of the GPE.
Then, we compare the GPE and the filament PDE of
Eqs.~(\ref{vr_eq3})--(\ref{vr_eq4})
for both stable and 
unstable modes predicted by the spectra. Our dynamical simulations also cover a wide range of trapping frequencies and chemical potentials.
In the figures showing GPE simulation results, we have plotted the VR as red points in 3D.
The vortical patterns are identified
through the methods of Refs.~\cite{foster,bisset}.
Additionally, we project both the VR position onto three back planes along with the 
density of the BEC, with contours of $0.2$, $0.4$, $0.6$, and $0.8$ of the
maximum density.

\begin{figure}
\includegraphics[width=0.45\columnwidth]{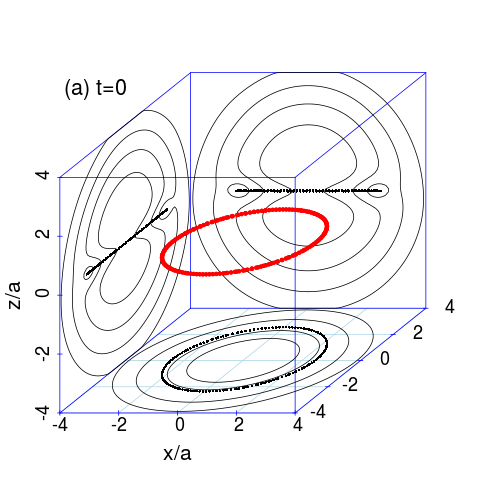}
\includegraphics[width=0.45\columnwidth]{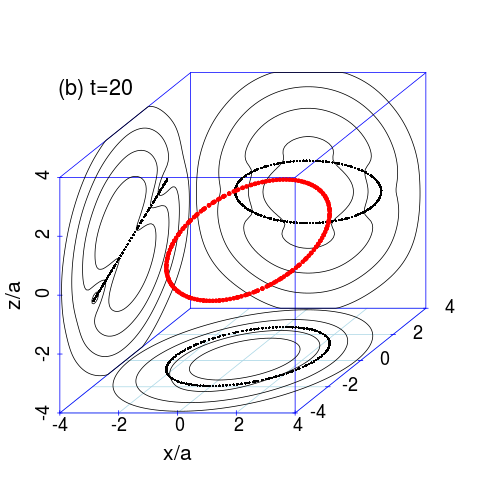}
\includegraphics[width=0.45\columnwidth]{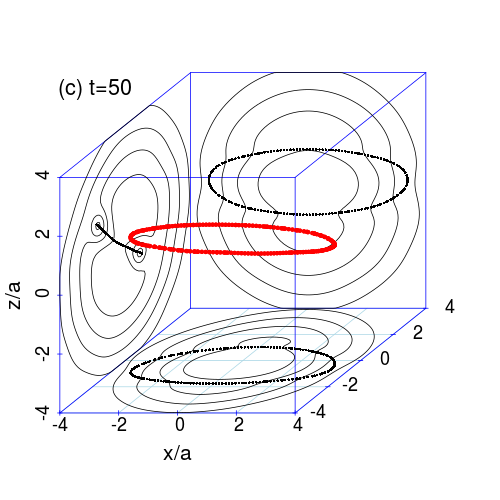}
\includegraphics[width=0.45\columnwidth]{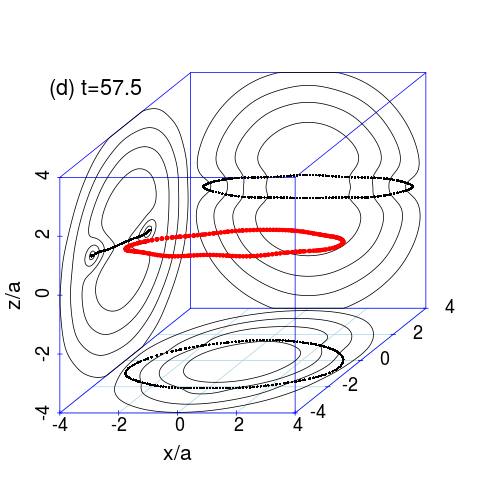}
\caption{Snapshots for $\lambda=0.95$, $\mu$=12, and $m=1$ show at
times $0, 20, 50$, and $57.5$ in units of $1/\omega_r$. Notice that the ring does not break for this mode, rather it flips over. 
The vortex ring leaves the BEC short after the last snapshot, presumably because the size of the condensate is small; see also Fig.~\ref{dyn2} at a larger chemical potential.
}
\label{dyn1}
\end{figure}

\begin{figure}
\includegraphics[width=0.45\columnwidth]{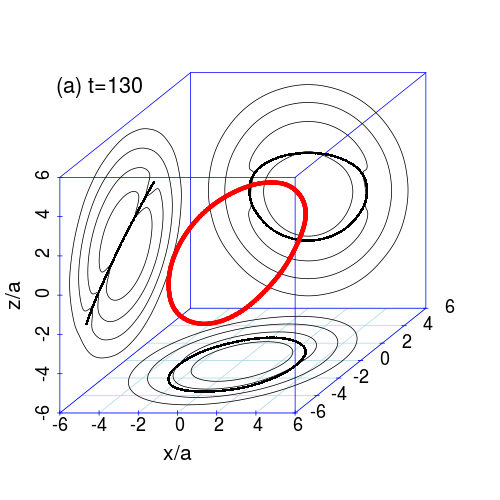}
\includegraphics[width=0.45\columnwidth]{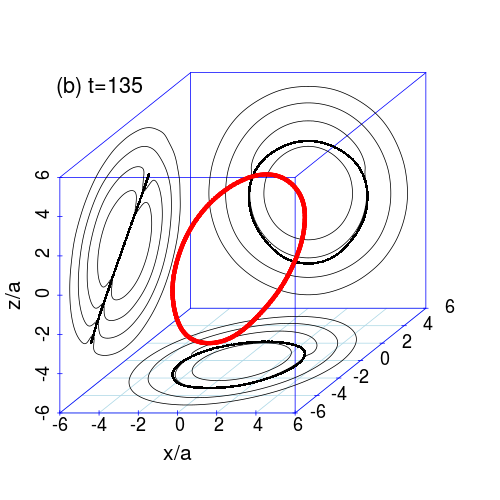}
\includegraphics[width=0.45\columnwidth]{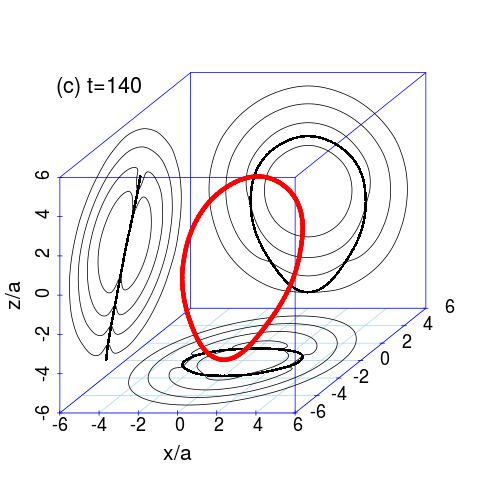}
\includegraphics[width=0.45\columnwidth]{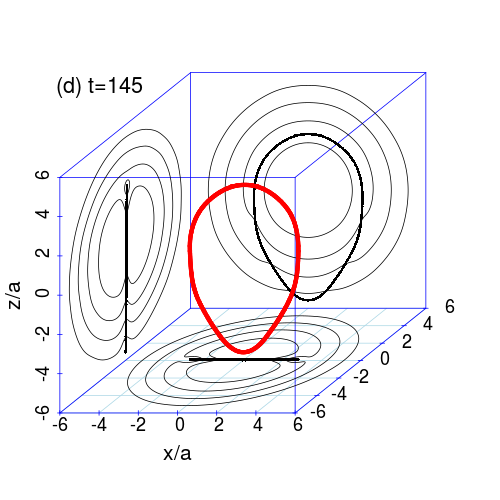}
\includegraphics[width=0.45\columnwidth]{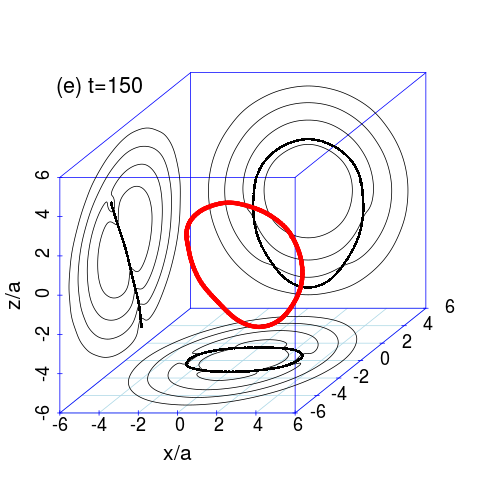}
\includegraphics[width=0.45\columnwidth]{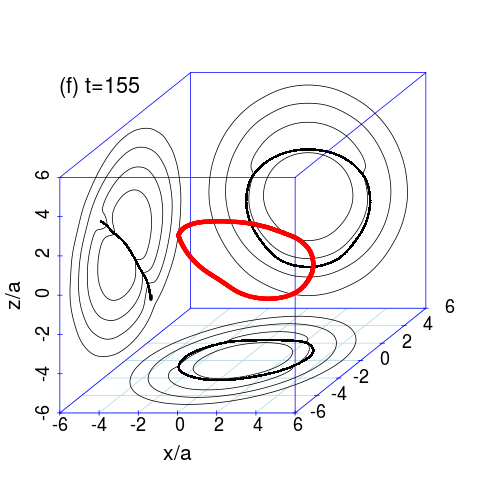}
\caption{Snapshots for $\lambda=0.95$, $\mu$=20, and $m=1$ shown at times: $130$, $135$, $140$, $145$, $150$, and $155$ in units of $1/\omega_r$.
The ring first flips over and then extends one side to the bottom of the BEC. However the ring does not break, rather it re-enters the BEC.}
\label{dyn2}
\end{figure}

The first pair of dynamical examples in Figs.~\ref{dyn1} and~\ref{dyn2}
concerns the instability of the mode with $m=1$ for $\lambda=0.95$.
We can see that indeed as we expect, this mode does not excite
undulations on top of the VR. Instead, it results in a dynamical rotation
of the ring within the (slightly
prolate) condensate that is combined with a growth of the radius
and partial deformation from a perfect cylindrical form. Nevertheless,
as can be seen in the figures, despite this instability the ring
remains robust and eventually its radius shrinks again, exhibiting a form
of (rotational) oscillatory dynamics.
For the same value of $\lambda=0.95$, but for larger values of $\mu$,
the ring is even more robust (in its nonlinear dynamics) in the
case examples we considered,
and takes a longer time before manifesting the flipping motion
that is observed as a result of the instability.

\begin{figure}
\includegraphics[width=0.45\columnwidth]{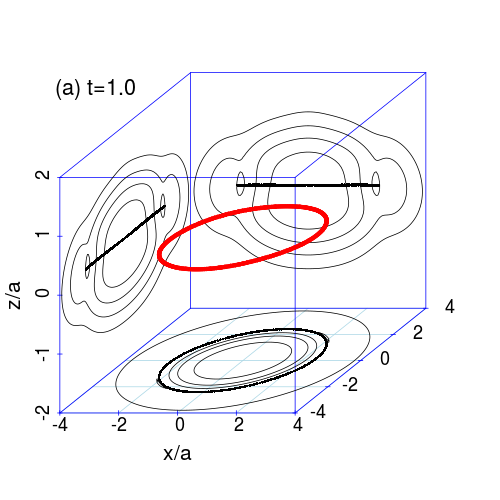}
\includegraphics[width=0.45\columnwidth]{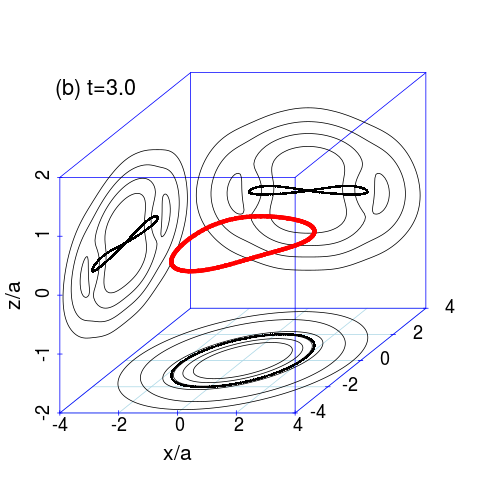}
\includegraphics[width=0.45\columnwidth]{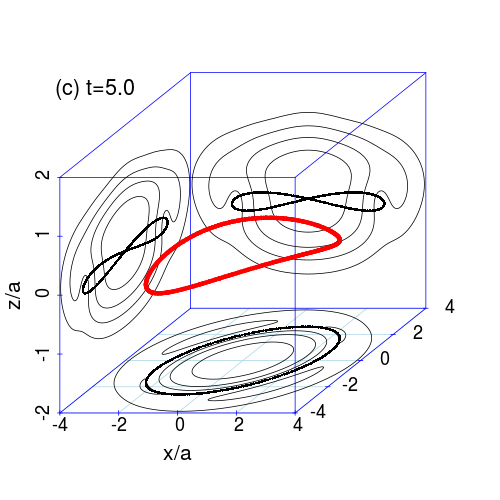}
\includegraphics[width=0.45\columnwidth]{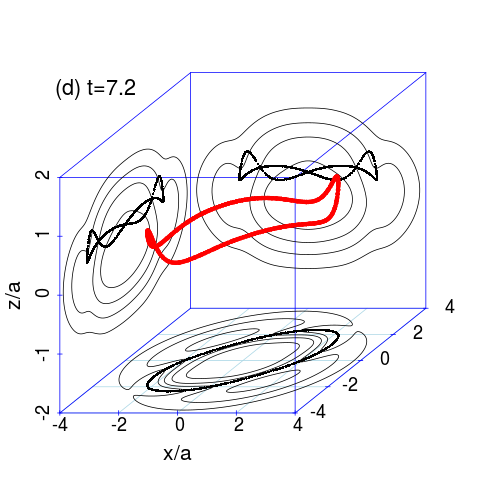}
\caption{Snapshots of the VR for $\lambda=3$, $\mu=12$, and an $m=2$
  mode perturbation shown at
times: $1.0$, $3.0$, $5.0$, and $7.2$ in units of $1/\omega_r$.
This shows the quick death of the vortex ring by the $m=2$ mode as the ring
just stretches horizontally out of the BEC.
}
\label{dyn3}
\end{figure}

\begin{figure}
\includegraphics[width=0.45\columnwidth]{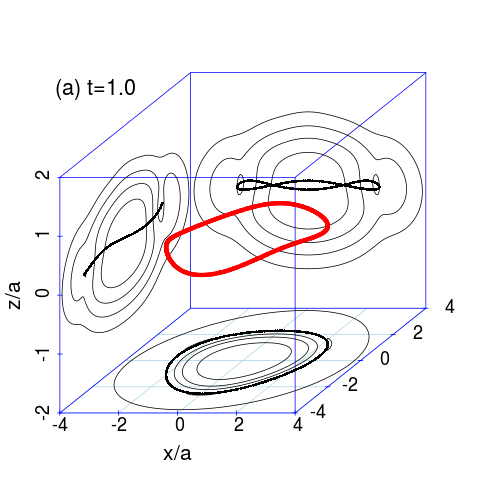}
\includegraphics[width=0.45\columnwidth]{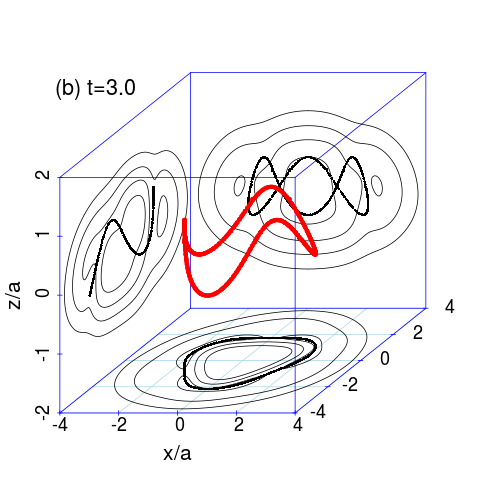}
\includegraphics[width=0.45\columnwidth]{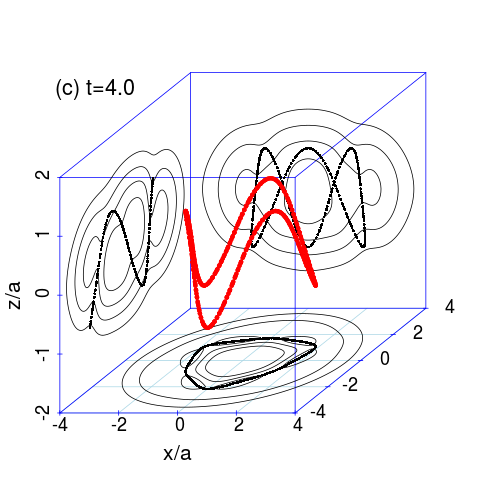}
\includegraphics[width=0.45\columnwidth]{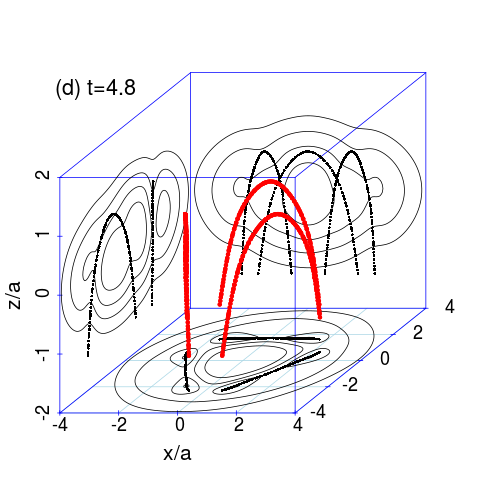}
\caption{Snapshots for $\lambda=3$, $\mu=12$, and an $m=3$ instability shown at
times: $1.0$, $3.0$, $4.0$, and $4.8$ in units of $1/\omega_r$. 
This shows the quick death of the vortex ring by the $m=3$ mode as the ring
breaks vertically out of the BEC.
}
\label{dyn4}
\end{figure}

We now turn to the unstable dynamics of the case with $\lambda=3$. Here,
in principle, given the finite $\mu$ nature of the computation,
{\it both} the $m=2$ and the $m=3$ modes have manifested their dynamical
instability.
The former is explored in the case of Fig.~\ref{dyn3},
while the latter is shown in Fig.~\ref{dyn4} [recall that the latter
  has a weaker instability growth rate and in fact would be completely
  stabilized for this $\lambda$ in the $\mu \rightarrow \infty$ limit].
In the case of Fig.~\ref{dyn3}, the quadrupolar nature of the
$m=2$ destabilizing mode is clearly evident both in the ring
dynamics, and in its corresponding planar projections shown in the
figures. Similarly, for the $m=3$ case, the hexapolar structure
can definitively be discerned in the top panels of Fig.~\ref{dyn4},
prior to the eventual breakup of the VR.

\begin{figure}
\includegraphics[width=0.45\columnwidth]{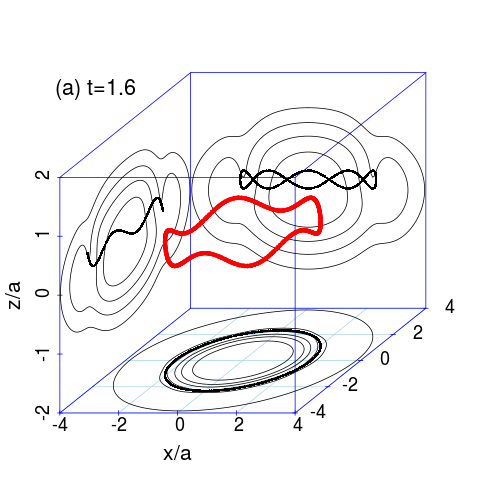}
\includegraphics[width=0.45\columnwidth]{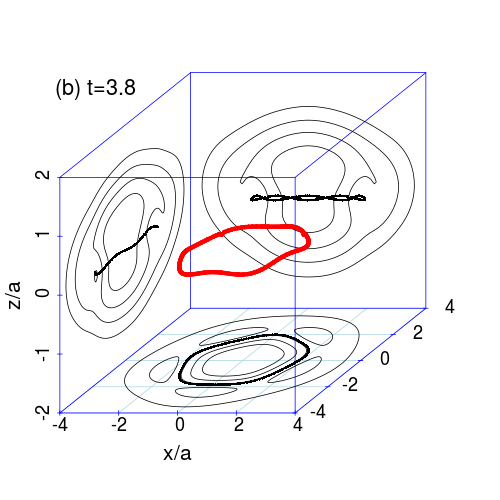}
\includegraphics[width=0.45\columnwidth]{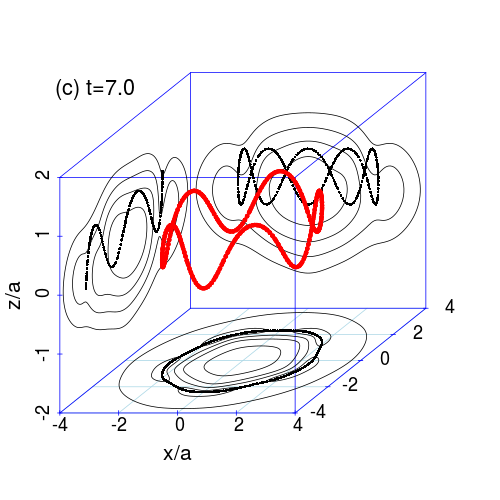}
\includegraphics[width=0.45\columnwidth]{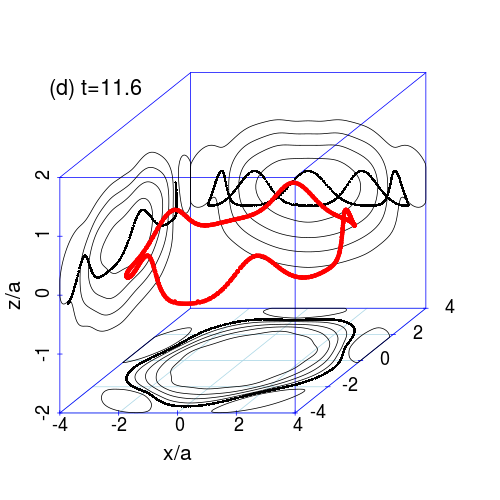}
\caption{Snapshots for $\lambda=3$, $\mu=12$, and $m=5$ shown at times: $1.6$,
  $3.8$, $7.0$, and $11.6$ in units of $1/\omega_r$.
  The mode is not unstable, yet
  it slowly evolves until it leaves the BEC, making
  for a long death of a vortex ring.}
\label{dyn5}
\end{figure}

An example of an excited stable $m=5$ mode for the case of $\lambda=3$
is shown in Fig.~\ref{dyn5}. The relevant mode is indeed excited
and persists for a long series of oscillations. However, eventually,
the small projection to other modes
of the (imperfect) initial data ``takes over''
manifesting the instability of the modes with $m=2$ and
$m=3$. This serves as a warning that while these higher $m$
modes may be stable, the ``contamination'' of the initial
data in practical situations with a small component in the
unstable modes will eventually lead to an instability,
even though this manifestation takes longer in this setting.

\begin{figure}
\includegraphics[width=0.45\columnwidth]{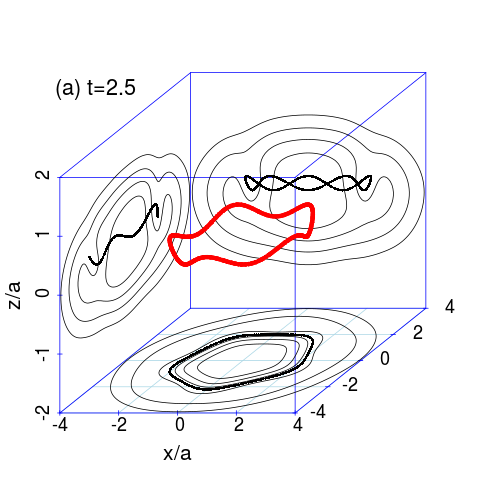}
\includegraphics[width=0.45\columnwidth]{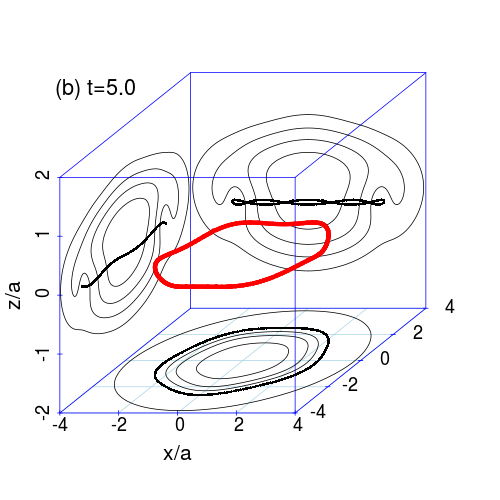}
\includegraphics[width=0.45\columnwidth]{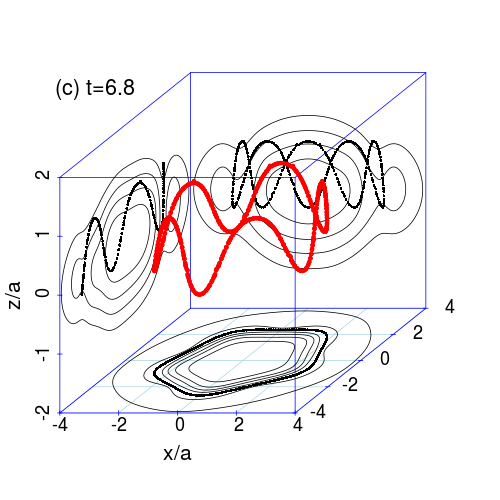}
\includegraphics[width=0.45\columnwidth]{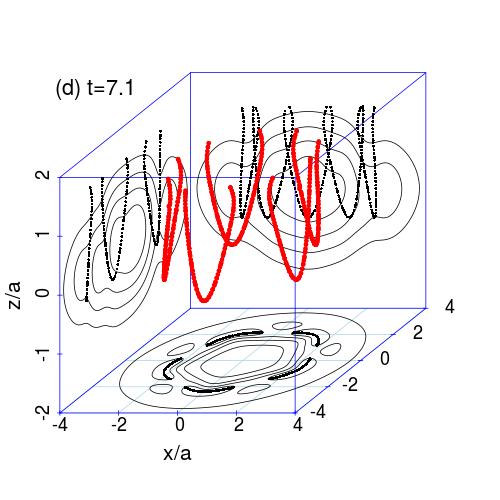}
\caption{Snapshots for $\lambda=3.1$, $\mu=12$, and $m=5$ shown at
times: $2.5$, $5.0$, $6.8$, and $7.1$ in units of $1/\omega_r$.
Here the ring quickly stretches and breaks vertically.
}
\label{dyn6}
\end{figure}

Lastly, as $\lambda$ becomes higher, as in Fig.~\ref{dyn6},
even the stable modes, such as the $m=5$, end up leading
to a relatively quick ``death'' of the VR. This is because
the unstable modes to which the initial data has some nontrivial
projection now have substantial growth rates that take over quickly
the relevant dynamical evolution, eventually leading to this
outcome. Fig.~\ref{dyn6} reports such an example for the
case of $\lambda=3.1$. Contrasting Figs.~\ref{dyn5} and \ref{dyn6} shows how only a seemingly slight
change in the geometry, for the same value of $\mu=12$,
alters the dynamics appreciably.
While Fig.~\ref{dyn6} shows a quick death for a vortex ring, 
Fig.~\ref{dyn5} shows a more protracted non-perturbative dance towards the vortex ring's demise.

Finally, we directly compare the filament PDE and the full
GPE dynamics of evolving the vortex ring, in order to get
a sense not only of the spectral but also of the dynamical
value of Eqs.~(\ref{vr_eq3})-(\ref{vr_eq4}). 
We first benchmark each method against their respective spectra for a stable oscillatory mode and then compare them for two selected unstable dynamical
scenarios
in the Thomas-Fermi limit. Recall from the spectra that the agreement of the two methods is expected to become progressively better
for all modes in the large chemical 
potential limit.

As a basis for comparison we have evolved both systems for the stable case of $\lambda=1.5$ and $\mu=20$ with a small vertical displacement of $0.1$ 
which corresponds to the $m=0$ mode. The motion of the vortex ring is an orbit around its (stable) equilibrium
radius. We have measured the frequencies of this motion from the two methods and they are in very good agreement with their respective
spectra. Since this scenario effectively involves a
simple harmonic motion around the respective equilibria,
we shall not discuss this further here. In what follows,
we compare the more interesting cases of the two dynamics
directly at a large chemical potential $\mu=40$.

\begin{figure*}
\includegraphics[width=0.45\columnwidth]{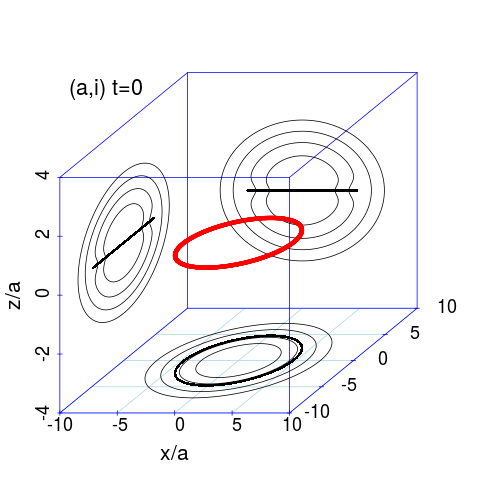}
\includegraphics[width=0.45\columnwidth]{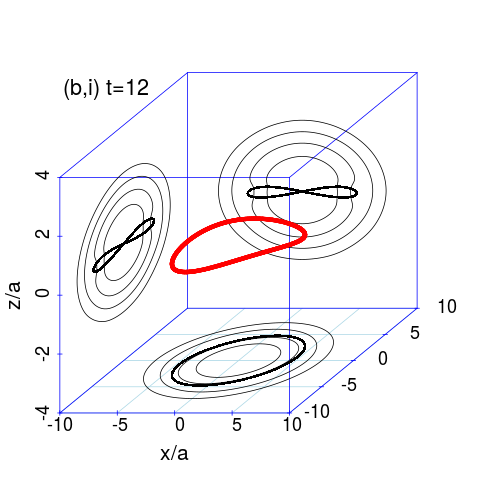}
\includegraphics[width=0.45\columnwidth]{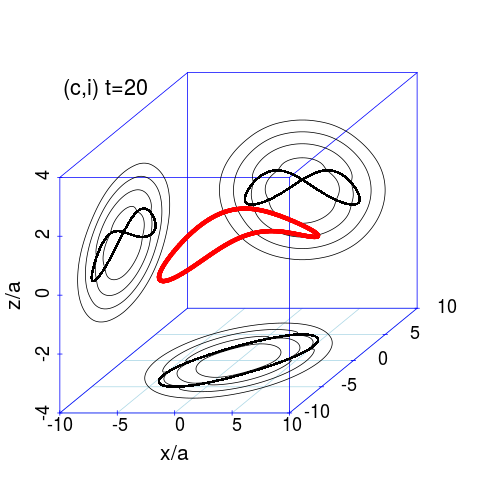}
\includegraphics[width=0.45\columnwidth]{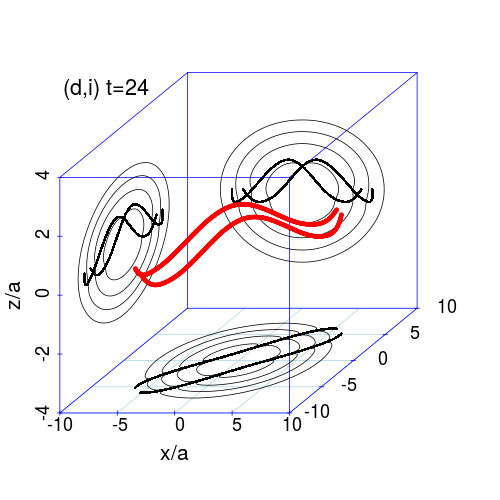}
\includegraphics[width=0.45\columnwidth]{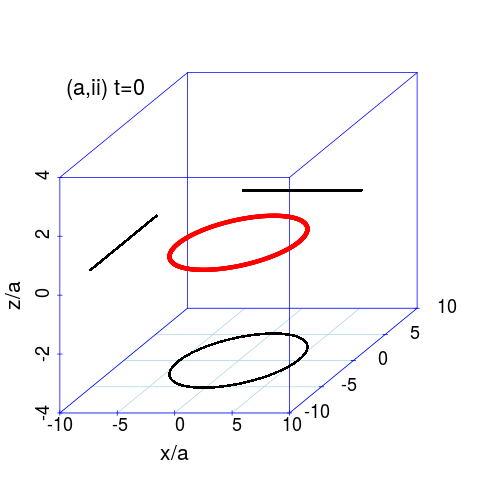}
\includegraphics[width=0.45\columnwidth]{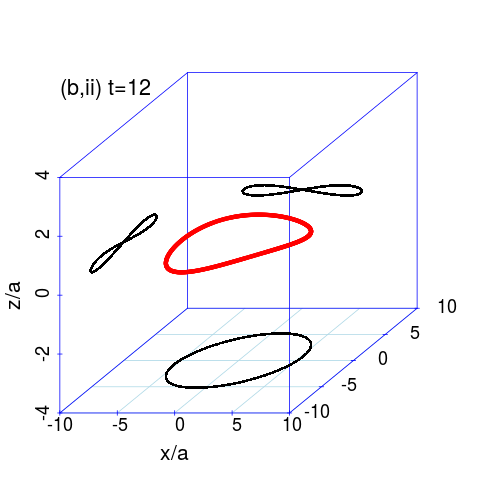}
\includegraphics[width=0.45\columnwidth]{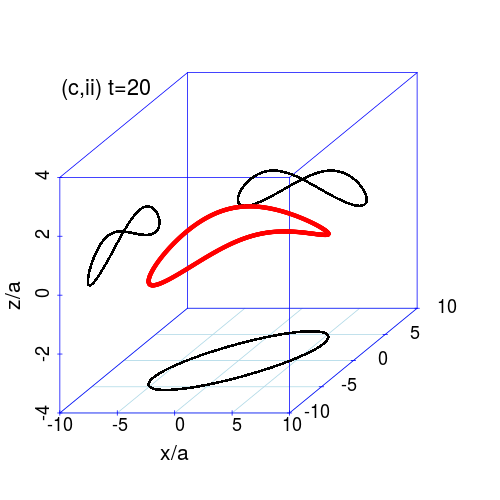}
\includegraphics[width=0.45\columnwidth]{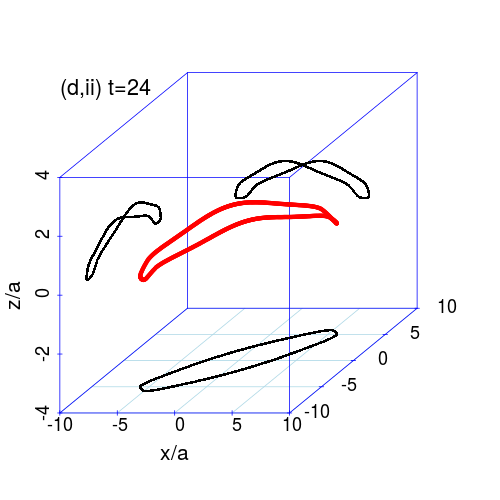}
\caption{
  Comparison of the filament PDEs (bottom panels) and the 3D GPE
  (top panels) at $\mu=40$ in the Thomas-Fermi limit. In this case, the unstable mode $m=2$ at $\lambda=3$ is excited. 
The comparison is presented at times 0, 12, 20, and 24 in units of $1/\omega_r$.
}
\label{m2}
\end{figure*}

\begin{figure*}
\includegraphics[width=0.45\columnwidth]{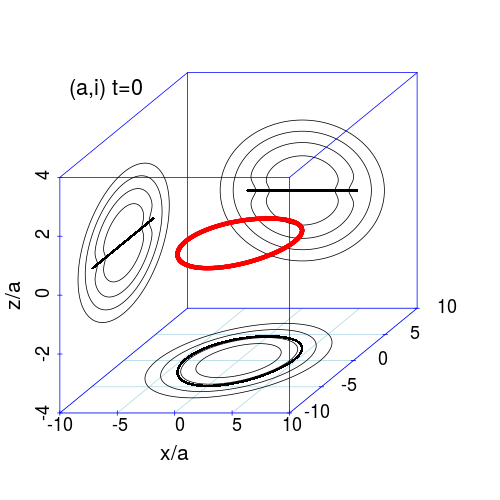}
\includegraphics[width=0.45\columnwidth]{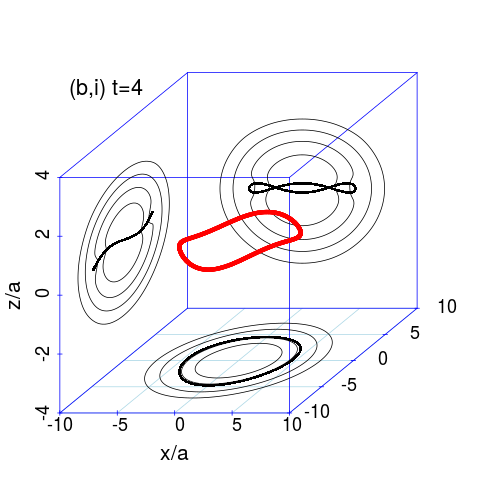}
\includegraphics[width=0.45\columnwidth]{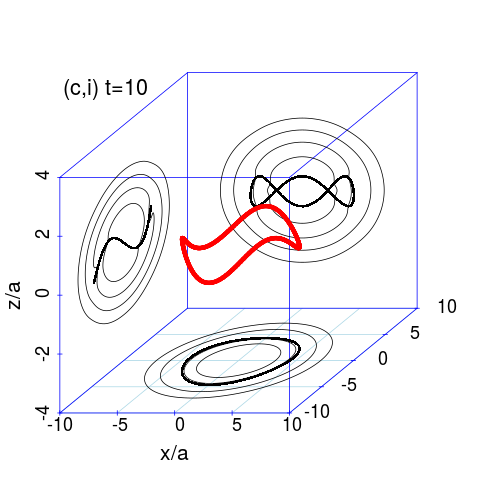}
\includegraphics[width=0.45\columnwidth]{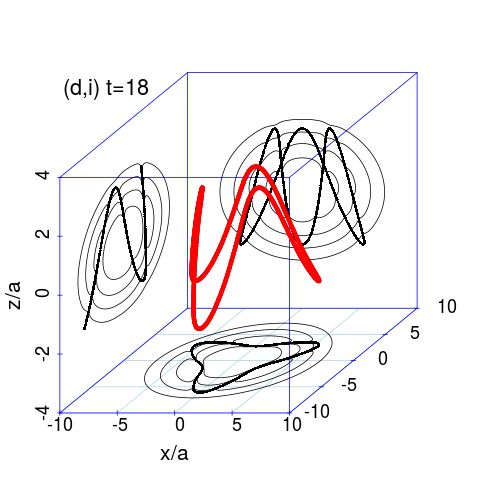}
\includegraphics[width=0.45\columnwidth]{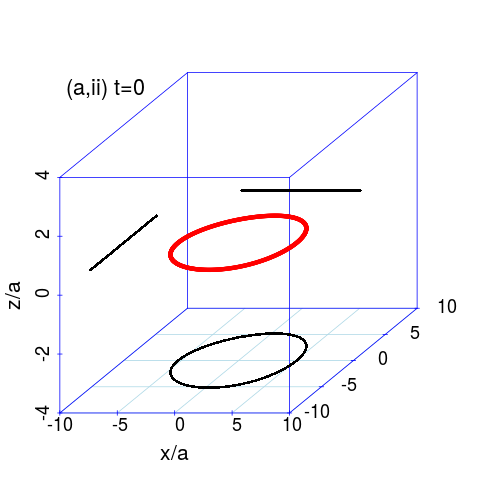}
\includegraphics[width=0.45\columnwidth]{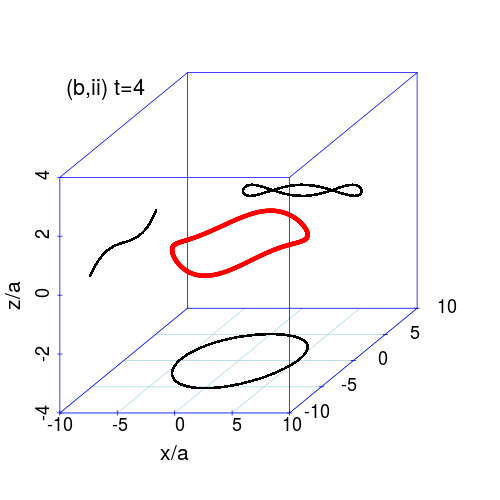}
\includegraphics[width=0.45\columnwidth]{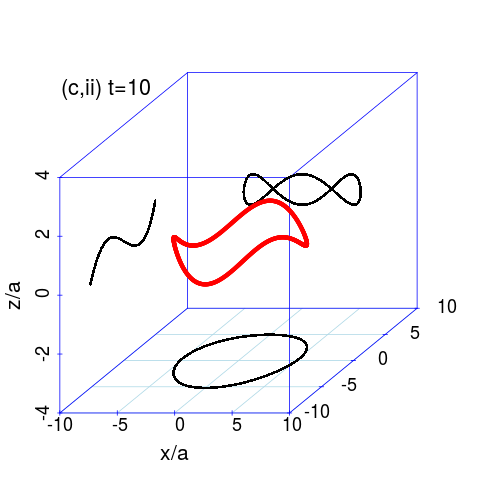}
\includegraphics[width=0.45\columnwidth]{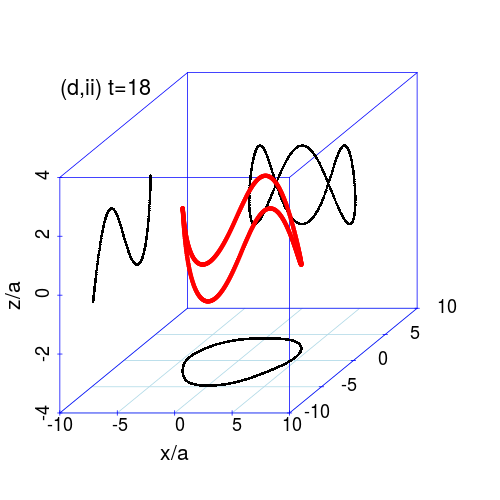}
\caption{
  Same as the previous figure, but for a case
  example where the unstable mode $m=3$ at $\lambda=3$ is excited. 
The comparison is presented at times 0, 4, 10, and 18 in units of $1/\omega_r$.
}
\label{m3}
\end{figure*}

We have again selected the $m=2$ and $m=3$ unstable modes at $\lambda=3$, such that this will allow the ring to deform rather than keeping the 
circular shape as in the $m=0$ mode case. The results for $m=2$ and $m=3$ are shown in Figs.~\ref{m2} and \ref{m3}, respectively. There are two important 
observations. One is that the filament PDEs indeed allow the ring to deform and break when an unstable mode is perturbed or excited, as is the case
in both of these scenarios. In contrast, the ODEs of 
Eq.~(\ref{vr_eq5}) do not allow the ring to depart from the perfect circular shape. Secondly, the filament PDEs are able to follow closely the 
dynamics of the GPE, in both the deformation patterns and the time scales before the ring breaks. While this may be natural to expect for short instability
times from our computed spectra, it is certainly far less obvious
for the full nonlinear evolution of Figs.~\ref{m2} and \ref{m3}.

\section{Conclusions \& Future Challenges}
\label{cc}

In the present work we have revisited the context of a vortical
ring filament, utilizing the Lagrangian and Hamiltonian formulation
thereof in order to go beyond the linearized (or simply time-dependent
ODE) formulation for the evolution of the ring. More concretely, we have
explored the full PDE evolution of the radius $R$ and vertical position
$Z$ of the vortex ring as a function of the azimuthal coordinate $\phi$ and
time. This
has allowed us to retrieve the  linearization results
from Ref.~\cite{horng}, which importantly have now been tested as a function
of both the nonlinearity strength (parametrized by the chemical
potential $\mu$) and of the trap geometry (parametrized by the
ratio $\lambda$ of longitudinal over transverse confinement). 
Indeed, the qualitative picture emerging from this analysis
was in excellent agreement with the numerical results and the
quantitative tendency of the latter to the former
as $\mu \rightarrow \infty$ was detailed. In prolate
condensates, the rings are unstable because they can start rotating
(an instability associated with the $m=1$ mode),
employing the freedom allowed in the vertical direction in such a case.
On the other hand, as the BEC transitions from a prolate to a weakly oblate
geometry, then it becomes stabilized (the spherical case is a special,
marginal example of this stabilization). However, if things become
too oblate (i.e., a ratio of 2:1 of the radial to transverse confinement
radius), then unstable modes, starting with the quadrupolar one
(and progressively higher ones as the anisotropy increases) become
unstable. As a result of the corresponding dynamical evolution, the
ring will break into an array of filaments, $2$ when the quadrupolar
mode is unstable, $3$ when it is the hexapolar, $5$ corresponding
to $m=5$ and so on. From the quantitative perspective, we find that
the analytically available spectrum very accurately reflects
the instability transition occurring at $\lambda=1$ and the associated
$m=1$ mode. However, it only captures higher modes (such as the $m=2$
going unstable when $\lambda=2$, the $m=3$ going unstable at $\lambda=3$
etc.) progressively more accurately in the limit of $\mu \rightarrow
\infty$. In these cases, the smaller the value of the chemical
potential, the more narrow the interval of ring stability becomes.

Naturally, these results pave the way for further developments as
regards the recently more and more accessible both experiment and
also measurement-wise vortical filaments. In particular, the formulation
of~\cite{ruban1,fetter1} can be extended to vortex lines and their
near-equilibrium linear, as well as highly nonlinear dynamics.
It would be particularly interesting to explore that case and
compare it systematically with the case of the ring, bearing in mind
also the destabilization of the ring into filamentary structures
in highly oblate condensates. This would create a fairly complete
picture of single filaments. Going beyond the single filament,
as is especially relevant in some experimental settings both
in superfluid Helium~\cite{wacks}, but also in BECs, this formulation,
along with that of Biot-Savart-based vortex interactions can lend
itself to the formulation of Hamiltonians and Lagrangians for the
case of multiple vortical filaments. This would allow us to capture
not only intriguing orbits involving multiple rings, such as
leapfrogging ones~\cite{caplan} (and their variants arising in a
trap~\cite{wang2}), but also the role that additional rings may play
in affecting the stability of a single ring, that was systematically
portrayed herein. Such studies are currently in progress and will be
reported in future publications.

\acknowledgments 
The authors are grateful to Prof. R. Carretero
for constructive comments on the manuscript.
W.W.~acknowledges support from the Swedish Research Council Grant 
No.~642-2013-7837 and Goran Gustafsson Foundation for Research in
Natural Sciences and Medicine.
P.G.K.~gratefully acknowledges the support of
NSF-PHY-1602994, as well as from  the Greek Diaspora
Fellowship Program. 
C. T. acknowledges support from the Advanced Simulation, and Computing and
LANL, which is operated by LANS, LLC for the NNSA of the
U.S. DOE under Contract No. DE-AC52-06NA25396.


\begin{thebibliography}{99}

\bibitem{saffman}
P.G. Saffman,
{\it Vortex Dynamics}
(Cambridge University Press, Cambridge, 1992).


\bibitem{Pismen1999}
L.M. Pismen,
{\it Vortices in Nonlinear Fields}
(Clarendon, Oxford, 1999).

\bibitem{donnelly}
R.J. Donnelly,
{\it {Q}uantized {V}ortices in {H}elium {II}}
(Cambridge University Press, Cambridge, 1991).

\bibitem{Rayfield64}
G.W. Rayfield and F.~Reif,
Phys. Rev. {\bf 136}, A1194--A1208 (1964).

\bibitem{Gamota73}
G.~Gamota,
Phys. Rev. Lett. {\bf 31}, 517--520 (1973).

\bibitem{book1} C. J. Pethick and H. Smith,
{\it Bose-Einstein Condensation in Dilute Gases} 
(Cambridge University Press, Cambridge, 2008).

\bibitem{book2} L. P. Pitaevskii and S. Stringari,
{\it Bose-Einstein Condensation} (Oxford University Press, Oxford, 2003).

\bibitem{dumitr1}
V. S. Bagnato, D. J. Frantzeskakis, P. G. Kevrekidis, B. A. Malomed, and
D. Mihalache,
Rom. Rep.
Phys. \textbf{67}, 5 (2015).

\bibitem{dumitr2}
D. Mihalache,
Rom. J. Phys. \textbf{59}, 295
(2014).

\bibitem{emergent}
P. G. Kevrekidis, D. J. Frantzeskakis, and R. Carretero-Gonz{\'a}lez (eds.),
{\it Emergent Nonlinear Phenomena in Bose-Einstein Condensates.
Theory and Experiment} (Springer-Verlag, Berlin, 2008);
R. Carretero-Gonz{\'a}lez, D. J. Frantzeskakis, and P. G. Kevrekidis, Nonlinearity {\bf 21}, R139 (2008).

\bibitem{komineas_rev} S. Komineas,
Eur. Phys. J.- Spec. Topics {\bf 147} 133 (2007).

\bibitem{barenghi_rev} C.F. Barenghi, R.J. Donnelly,
  Fluid. Dyn. Res. {\bf 41}, 051401 (2009).

\bibitem{book_new} P. G. Kevrekidis, D. J. Frantzeskakis,
and R. Carretero-Gonz{\'a}lez, {\it The defocusing Nonlinear Schr{\"o}dinger Equation:
From Dark Solitons to Vortices and Vortex Rings} (SIAM, Philadelphia, 2015).

\bibitem{Anderson01}
B.P. Anderson, P.C. Haljan, C.A. Regal, D.L. Feder, L.A. Collins,
C.W. Clark, and E.A. Cornell,
Phys. Rev. Lett. {\bf 86}, 2926--2929 (2001).

\bibitem{Shomroni09}
I.~Shomroni, E.~Lahoud, S.~Levy, and J.~Steinhauer,
Nat. Phys. {\bf 5}, 193--197 (2009).


\bibitem{Ginsberg05}
N.~S. Ginsberg, J.~Brand, and L.V. Hau,
Phys. Rev. Lett. {\bf 94}, 040403 (2005).

\bibitem{sengstock}
C.~Becker, K.~Sengstock, P.~Schmelcher, P.G. Kevrekidis, and R.~Carretero-Gonz{\'a}lez,
New J. Phys. {\bf 15}, 113028 (2013).

\bibitem{bis1} R. N. Bisset, S. Serafini, E. Iseni, M. Barbiero, T. Bienaime, G. Lamporesi, G. Ferrari, F. Dalfovo,
  Phys. Rev. A {\bf 96}, 053605 (2017). 

\bibitem{bis2} F. Dalfovo, R. N. Bisset, C. Mordini, G. Lamporesi and G. Ferrari,
  arxiv:1804.03017 (2018)

\bibitem{lathrop1} G. P. Bewley, M. S. Paoletti,
  K. R. Sreenivasan, and D. P.
Lathrop,
Proc. Natl. Acad. Sci. U.S.A.
{\bf 105},
13707 (2008).

\bibitem{lathrop2} E. Fonda, D. P. Meichle, N. T. Ouellette,
  S. Hormoz, and D. P. Lathrop,
Proc. Natl. Acad. Sci.
U.S.A. {\bf 111}, 4707 (2014).
  

\bibitem{wacks} D.H. Wacks, A.W. Baggaley, C.F. Barenghi,
  Phys. Rev. B {\bf 90}, 224514 (2014).

\bibitem{wang2} Wenlong Wang, R. N. Bisset, C. Ticknor, R. Carretero-Gonzalez, D. J. Frantzeskakis, L. A. Collins, P. G. Kevrekidis,
  Phys. Rev. A {\bf 95}, 043638 (2017).
  

\bibitem{bis3} S. Serafini, L. Galantucci, E. Iseni, T. Bienaim{\'e},
  R. N. Bisset, C.F. Barenghi, F. Dalfovo, G. Lamporesi and G. Ferrari,
  Phys. Rev. X {\bf 7}, 021031 (2017).

  

  
  \bibitem{Tsatsos2016}
M.~C.~Tsatsos, P.~E.~S.~Tavares, A.~Cidrim, A.~R.~Fritsch, M.~A.~Caracanhas,  F.~Ednilson, A.~dos Santos, C.~Barenghi, and V.~S.~Bagnato,
Phys. Rep. {\bf 622}, 1 (2016).

\bibitem{Navon2016}
N.~Navon,  A.~L.~Gaunt, R.~P.~Smith, Z.~Hadzibabic,
Nature {\bf 539}, 73 (2016).

\bibitem{zwierlein} M.J.-H. Ku, B. Mukherjee, T. Yefsah, and M.W. Zwierlein,
  Phys. Rev. Lett. {\bf 116}, 045304 (2016).


\bibitem{ruban1} V. P. Ruban,
Phys. Rev. E {\bf 64}, 036305 (2001). 

\bibitem{fetter1} A.A. Svidzinsky and A.L. Fetter,
Phys. Rev. A {\bf 62}, 063617 (2000).

\bibitem{horng}  T.-L. Horng, S.-C. Gou, T.-C. Lin,
  Phys. Rev. A {\bf 74}, 041603(R) (2006).
  
\bibitem{ruban2} V.P. Ruban, JETP Lett. {\bf 106}, 223 (2017).

\bibitem{wenlongdss} 
W. Wang, P. G. Kevrekidis, R. Carretero-Gonz{\'a}lez, and D. J. Frantzeskakis,
Phys. Rev. A {\bf 93}, 023630 (2016).

\bibitem{tickold} L.-C. Crasovan, V.M. P{\'e}rez-Garc{\'i}a,
  I. Danaila, D. Mihalache,
  and L. Torner,
  Phys. Rev. A {\bf 70},
033605 (2004);
R. N. Bisset, W. Wang, C. Ticknor, R. Carretero-Gonz{\'a}lez,
D. J. Frantzeskakis, L. A. Collins, and P. G. Kevrekidis,
Phys. Rev. A {\bf 92}, 043601 (2015). 
  
\bibitem{ai1} P. G. Kevrekidis, W. Wang, R. Carretero-Gonz{\'a}lez, and D. J. Frantzeskakis,
Phys. Rev. Lett. {\bf 118}, 244101 (2017).

 
 
\bibitem{foster} C. J. Foster, P. B. Blakie, and M. J. Davis, Phys. Rev. A, {\bf81}, 023623 (2010).

\bibitem{bisset} R. N. Bisset, S. Serafini,E. Iseni, M.Barbiero, T. Bienaim\'e, G. Lamporesi, G. Ferrari, and F. Dalfovo,
 Phys. Rev. A {\bf 96},053605 (2017).

 
\bibitem{caplan} R. M. Caplan, J. D. Talley, R. Carretero-Gonz{\'a}lez,
  P.G. Kevrekidis, Phys. Fluids {\bf 26}, 097101 (2014).

\end{thebibliography}
\end{document}